\documentclass{pas}

\usepackage{amsmath}
\usepackage{multirow}

\begin{document}

\lefttitle{M dwarfs quasi-periodic pulsations at a time resolution of 1 s}
\righttitle{A. Panferov, G. Beskin, S. Karpov {\it et al.}}

\jnlPage{1}{9}
\jnlDoiYr{2024}
\doival{10.1017/pasa.xxxx.xx}

\articletitt{Research Paper}

\title{M dwarfs quasi-periodic pulsations at a time resolution of 1 s}

\author{\sn{Alexander} \gn{A.~Panferov}$^{1,2}$, \sn{Grigory} \gn{Beskin}$^{1}$, \sn{Sergey} \gn{Karpov}$^{3}$, \sn{Vladimir} \gn{Plokhotnichenko}$^{1}$ and \sn{Olga} \gn{Maryeva}$^{4}$}

\affil{
$^1$Special Astrophysical Observatory of  Russian Academy of Sciences, N. Arkhyz 369167, Russia \\
$^2$Togliatti State University, Togliatti 445667, Russia \\
$^3$Institute of Physics of the Czech Academy of Sciences, CZ-182 21 Prague 8, Czech Republic \\ 
$^4$Astronomical Institute of the Czech Acad. of Sci., Fricova 298, 25165 Ondrejov, Czech Republic
}

\corresp{Grigory Beskin, Email: beskin@sao.ru}

\citeauth{Beskin G., Karpov S., Plokhotnichenko V., Maryeva O., and Panferov A. A., M dwarfs quasi-periodic pulsations at a time resolution of 1 s. {\it Publications of the Astronomical Society of Australia} {\bf 00}, 1--9. https://doi.org/10.1017/pasa.xxxx.xx}

\history{(Received xx xx xxxx; revised xx xx xxxx; accepted xx xx xxxx)}

\begin{abstract}
Quasi-periodic pulsations (QPPs) of Sun and stars are challenging for stellar flare models. The white light stellar QPPs in the periodicity region of tens of second are unexplored yet. 
On the basis of observations with the 6-m telescope BTA in $U$-band of flaring dM-stars EV~Lac, Wolf~359, Wolf~424, V577~Mon and UV~Ceti we found 13 new QPPs. This composes 30\% occurrence among 44 worked flares. These QPPs were found to have periods ranging from 6 to 107 seconds and were detected using both Fourier transform and empirical mode decomposition methods. The observed QPPs were categorized by the evolution of their oscillation envelope and fractional flux amplitudes. There are shown the statistically significant correlations of the QPP period with the duration, the equivalent duration and the amplitude of a flare, and the correlation between the QPP amplitude and flare amplitude.
\end{abstract}


\maketitle

\section{Introduction}

Quasi-periodic pulsations (QPPs) appear in the light curves of flares of Sun as well as stars \citep{Zim21}. The number of identified stellar QPPs is quite small, 213 on 2021 year \citep{Zim21}, that is in particular why they are of interest. Most of them are the optical QPPs, on the one hand, and of red dwarfs, on the other hand. QPPs could be used to obtain properties of flare regions and are a problem for flare models yet. QPPs are identified in stellar flares in the spectral bands from radio to X-ray. In the optical band they are searched mainly in the data of the observatories $\it Kepler$ and TESS. The periods of QPPs from these data are much larger (of tens of minutes) than from the data of high cadence. So, the high cadence observations with the 6-m telescope BTA (this paper) and the space-borne observatory GALEX \citep{Doy18} revealed the periods of tens of seconds. However, to this day the most optical QPPs are of periods larger than $\sim 10$ minutes \citep{HM22}, while there were observed only a few optical QPPs with cadence of 1~s or higher \citep{Doy22}, and the domain of periods of tens of second is unexplored. On the other hand, observations of solar flares which are usually weaker (of the energies $\ll 10^{32}$~erg) than stellar ones are scarce of detected optical QPPs. Therefore observations of stellar QPPs in optics are important for disclosure of the mechanisms under the QPPs. 

In this paper, we analyse the sample of 44 flares of M dwarfs for QPPs. The observation are presented in the Section~\ref{Ob}. And the Section~\ref{M} deals with the analysis of light curves using Fourier and the empirical mode decomposition methods. The results of the search for QPPs are discussed in the Section~\ref{R}. The Section~\ref{C} is conclusive.

\section{Observations}
\label{Ob}
\begin{figure}[hbt!]  
\includegraphics[width=0.8\linewidth]{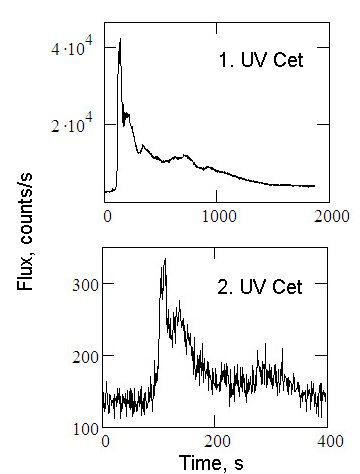}
\centering
\caption{The light curves in flares of M dwarf UV~Ceti in $U$-band observed with MANIA/BTA, where QPPs were found. The QPP number in Table~\ref{T1} precedes the star name on panels. The time zero point is at a flare beginning.}        
\label{LCa}
\end{figure}

\begin{figure}[hbt!]
\centering
\includegraphics[width=0.8\linewidth]{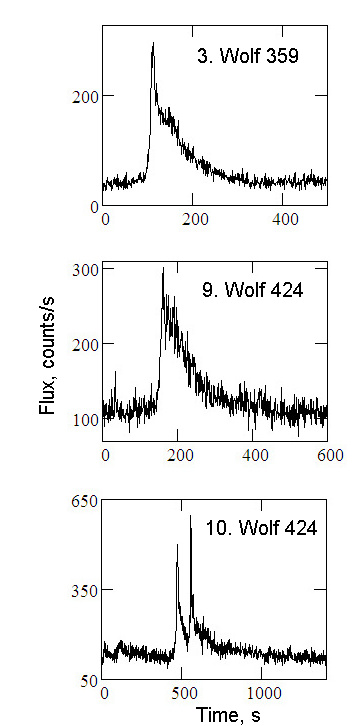}
\caption{The light curves in flares of M dwarfs Wolf~359 and Wolf~424 in $U$-band observed with MANIA/BTA, where QPPs were found. The QPP number in Table~\ref{T1} precedes the star name on panels. The time zero point is at a flare beginning.}
\label{LCb}
\end{figure}

\begin{figure*}
\centering
\includegraphics[width=1.0\linewidth]{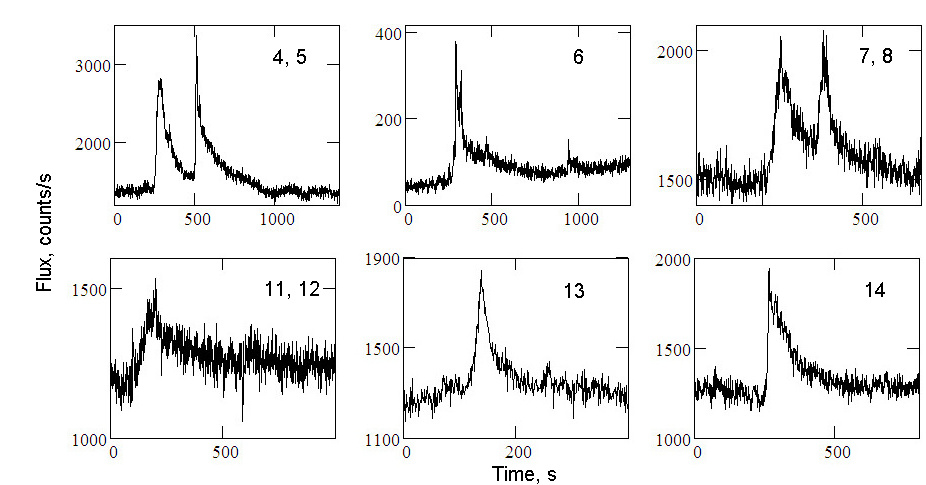}
\caption{The light curves in flares of M dwarf EV~Lac in $U$-band observed with MANIA/BTA, where QPPs were found. The QPP numbers in Table~\ref{T1} mark panels. The time zero point is at a flare beginning.}
\label{LCc}
\end{figure*}

We undertook 70~hours observational campaign with the MANIA complex \citep{Pl21} on the 6-m BTA telescope in $U$-band to search for fast variability of the nearest M dwarfs at microsecond time resolution. 
The complex consists of a MPPP multimode panoramic photometer-polarimeter \citep{mppp} based on a coordinate-sensitive detector \citep{psd} equipped with a set of photometric filters, dichroics and a double Wollaston prism, a time-tagging device for determining the moments of registration and coordinates of incoming photons, as well as a set of computers for accumulating and storing the data as photon lists that allow further analysis of variability on shortest possible time scales.
In these observations we found 157 flares of dM-stars EV~Lac, Wolf~359, Wolf~424, V577~Mon and UV~Ceti. 44 of them were inspected for QPPs. For these flares we used an effective time cadence of 1~s since periodicity shorter than 5~s was not identified. The light curves of the flares with detected QPPs are shown in Fig.~\ref{LCa}, \ref{LCb} and \ref{LCc}. The list of these QPPs with dates of observations and physical characteristics is given in Table~\ref{T1}. 

\section{Search for QPPs}
\label{M}
For the analysis of flare light curves on periodical signal we used two independent methods, the Fourier transform and the empirical mode decomposition (EMD),  \citep{Hu98,Hu08}. The latter applied to QPPs is extensively used \citep{Kol16,Kol18,Doy18,Ram21,Fl22}. EMD is adaptive to non-linear and non-stationary signals, with a period drift, which are QPPs, unlike the Fourier transform method rooted in the preset of harmonic modes for the decomposition. By the review of \citet{Br19} (which methods we adhere) the using more than one of methods increases detectability of QPPs otherwise missed due to fast decaying and non-stationarity of periods. Nevertheless, they also note that the probability of the missing of QPPs is quite high even using several methods. 

\begin{figure*}
\centering
\includegraphics[width=1.0\linewidth]{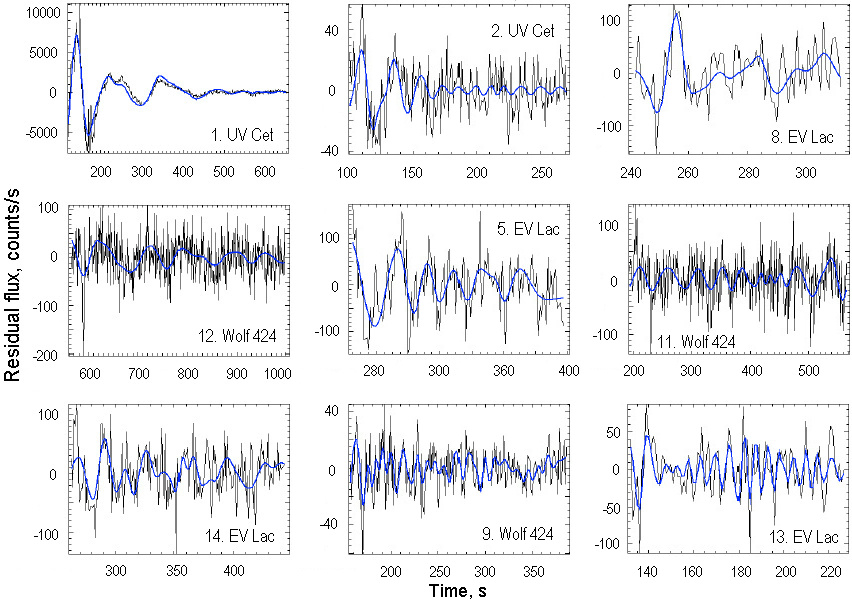}
\caption{The pulsation components of flares of M dwarfs with the superposed blue line of EMD-modes. The QPP number in Table~\ref{T1} precedes the star name on graphs. The time zero point is at a flare beginning. To highlight a differentiation the QPPs are grouped: the decaying ones into the left column, the transitional ones into the central column, and the decayless ones into the right column.}
\label{PLCa}
\end{figure*}

\begin{figure}[hbt!]
\centering
\includegraphics[width=0.7\linewidth]{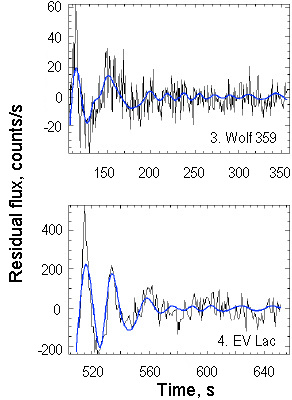}
\caption{The pulsation components of flares of M dwarfs with the superposed blue line of EMD-modes. The QPP number in Table~\ref{T1} precedes the star name on graphs. The time zero point is at a flare beginning.}
\label{PLCb}
\end{figure}

\begin{figure}[hbt!]
\centering
\includegraphics[width=0.7\linewidth]{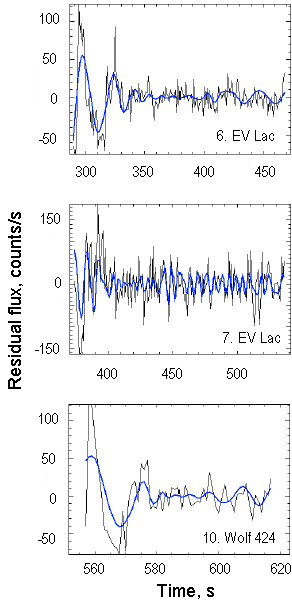}
\caption{The pulsation components of flares of M dwarfs with the superposed blue line of EMD-modes. The QPP number in Table~\ref{T1} precedes the star name on graphs. The time zero point is at a flare beginning.}
\label{PLCc}
\end{figure}

On pulsations was analysed the residual $I_{\rm p} (t)$ of the light curve $I(t)$ after subtraction of a flare trend. In the first approximation, the flare trend $I_{\rm fl\,0} (t)$ was prescribed as the superposition of a linear function and a Gaussian before the flare peak $t_{\rm peak}$ or an exponent after the flare peak, which $t_{\rm peak}$, the Gaussian half-width $t_{\rm in}$ and the exponential decay time $t_{\rm dec}$ were determined from the fit to the flare light curve section above the level 0.3 of maximum flare flux over the quiescent level $I_{\rm q} (t)$ (for star without flare). In this approximation, the pulsation component $I_{\rm p\,0} (t) = I(t) - I_{\rm fl\,0} (t)$ was decomposed in EMD-modes (called intrinsic mode functions) in all light curve sections $[t_{\rm peak}-t_{\rm in}, t_{\rm peak}+k t_{\rm dec}]$, where $k=1$, 2 $\ldots$ 5, with the number of data point no less than 31 in a section. With the sum $I_{\rm fl\,0.4}(t)$ of the slow EMD-modes, or the trend-modes, of the characteristic time scales longer than 0.4 the section length, we redefined the flare trend $I_{\rm fl} (t) = I_{\rm fl\,0} (t) + I_{\rm fl\,0.4} (t)$ and the pulsation component $I_{\rm p} (t) = I(t) - I_{\rm fl}(t)$, in the second approximation. So derived pulsation components of the light curves (in other words, the detrended light curves) are shown in Fig.~\ref{PLCa},  \ref{PLCb} and \ref{PLCc}.

The periods of possible pulsations in the flares were revealed by the Fourier spectrum of the pulsation component $I_{\rm p} (t)$, and their statistical significance were assessed by comparing the power $S(f_{\rm j})$ of a peak at the frequency $f_{\rm j}$ in the Fourier spectrum with the confidence level $P_\alpha(f)$, corresponding to $100 \alpha$~\% confidence probability at the given frequency $f_{\rm j}$. The confidence level was determined by the formula ${P_\alpha(f) = P(f) \ln(N_f/(1 - \alpha)) \left<S(f_{\rm j})/P(f_{\rm j})\right>}$ (\citet{Br19}, section 4.7), where $P(f)$ is the power-law fit to the spectral power $S(f_{\rm j})$, $<>$ means an averaging over the spectrum, $N_f$ is the number of values in the spectrum. The pulsations with power $S(f_{\rm j})$ exceeding the threshold level $P_\alpha(f_{\rm j})$ corresponding to $\alpha = 0.68$ (the $1\sigma$ level) were further analyzed on the EMD statistical significance. The analysis of pulsations was confined to the period interval 4 times the time cadence -- 0.4 times the light curve section length. 

Similarly to the recognition the signal over noise on the Fourier spectrum the pulsations represented by EMD-modes could be verified vs. the confidence probability level in the EMD spectrum, the dependency of the mode energy $E$ on the period $P$. Here, the mode energy $E$ is determined as a sum of squares of the instantaneous amplitudes of the mode signal, and its period $P$ is the period of peak of the mode in the wavelet spectrum. For a noise time series an EMD-mode energy has chi-squared statistics which could be indicated in the EMD spectrum by the confidence levels $P_\alpha(P)$ for the extracted above EMD-modes. The pulsations with EMD-modes exceeding the threshold level $P_\alpha(P_{\rm j})$ corresponding to $\alpha = 0.95$ (the $2\sigma$ level) were throughout selected. In that way the detected QPPs have at least $1\sigma$ (68\%) significance in the Fourier analysis and $2\sigma$ (95\%) significance in the EMD analysis. The period and its uncertainty of an EMD-mode were determined from the Gaussian fitted to the corresponding wavelet spectrum. 

\section{Results}
\label{R}
All BTA flares in Fig.~\ref{LCa}, \ref{LCb} and \ref{LCc} look like one of the classical flare profile \citep{Dav14} with such added morphology complexities as peak-bump, flat-top peak, spiked peak and QPPs. \citet{HM22} thoroughly studied occurrence of these complexities on a basis of 440 TESS flares of 20~s cadence. 13 QPPs of 1~s effective cadence are shown in Fig.~\ref{PLCa}, \ref{PLCb} and \ref{PLCc} (we note that QPPs~11 and 12 correspond to the same flare) where pulsating components are prominent above the detrended flare light curves at the levels at least of $1\sigma$ significance in the Fourier analysis and $2\sigma$ significance in the EMD analysis described above. They are modeled by the EMD-modes, shown by blue line, which periods $P_{\rm QPP}$ spread in the range 6--107~s, as seen from Table~\ref{T1} of the flares characteristics. One can also see in Fig.~\ref{PLCa}, \ref{PLCb} and \ref{PLCc}, the QPPs evolve in a flare, both in the period and amplitude, while the solar QPPs are mainly of narrow band periodicity. There are multiperiodical QPPs (1, 8, 9 and 14), which periods are multiple within their uncertainties. This is hard support to consider them as overtones and the standing waves as ingredient of underlaying mechanisms of these QPPs.   

\subsection{Occurrence and periodicities of QPPs}

The detected 13 QPPs of the 44 flares implies 30\% occurrence rate. This is much more than 3\% occurrence rate by \citet{Bal15}, deduced for the $\it Kepler$ sample obtained with 1~min cadence, 4\% occurrence rate by \citet{Pug16}, also for the $\it Kepler$ sample issued in the QPPs periods 5--93~min, 7\% occurrence rate by \citet{Ram21}, for the TESS sample obtained with 2~min cadence, also issued in the similar QPPs periods, 10--72~min, 5\% occurrence rate by \citet{Fl22} for the GALEX observations of UV Cet at 1~s cadence, and 4\% occurrence rate by \citet{HM22}, also for the TESS sample but obtained with 20~s cadence issued in the QPPs periods 3--25~minutes which are mostly less than 10~min. The range 6--107~s of the found here periods is very close to the periods 20--120~s by \citet{Doy18} for six dMe stars observed with GALEX at 1~s cadence. Our findings on the occurrence and periodicities of QPPs consist with the high occurrence of solar X-ray QPPs, changing from 7\% to 46\% for the C through X flare classes (which represents distribution of flares in energy), and with the period range $\sim 10$--40~s of the majority of solar QPPs \citep{Hay20}. Evidently, the boost of the stellar optical QPPs with periods of order of some tens of second became possible due to fast enough cadence, 1~s. By an analogy, the increase of cadence led \citet{HM22} to the detection of bulk QPPs under 10~min periodicity, previously weakly populated periodicity domain. However, with an eye on the solar QPPs (of high occurrence), it is not clear is the found here occurrence of the M dwarfs QPPs, in the periodicity domain of tens of second, so high due to either the high time resolution of observations or the high sensitivity of observations, which allowed to detect the QPPs in the low energy flares, of the energies mostly $\sim 10^{30}$--$10^{32}$~erg (see Table~\ref{T1}).

\begin{table*}
\centering
\caption{Characteristics of flares and QPPs: QPP index, star name, observation date, QPP period, flare duration, QPP amplitude, flare amplitude, flare magnitude in $U$ filter, flare equivalent duration, flare bolometric energy.}
\begin{tabular}{c c c r r r r r r r }
\hline
\hline
QPP & Star & Date & $P_{\rm QPP}$ & ${\rm \Delta}t_{\rm fl}$ & $A_{\rm QPP}$ & $A_{\rm fl}$ & ${\rm \Delta}m_{\rm U}$ & $W_{\rm t}$ & $E_{\rm fl}$ \\
& & & (s)&  (s)&  &  &  & (s)& ($10^{32}$~erg)\\
\hline
1 & UV Ceti & 2008-12-28 & \mbox{$70\pm19$} & 1817 & 0.1 & 14.4 & 3.0 & 4210 & 0.43\\
 & & & $107\pm18$ & & & & & & \\
2 & UV Ceti & 2009-01-01 & $24\pm7$ & 262 & 0.08 & 1.46 & 1.0 & 92 & 0.009\\
3 & Wolf 359& 2009-01-28 & $46\pm20$ & 260 & 0.1 & 6.17 & 2.1 & 317 & 0.038\\
4 & EV Lac & 2009-07-19 & $19\pm4$ & 440 & 0.07 & 1.47 & 1.0 & 119 & 0.63\\
5 & EV Lac & 2009-07-19 & $26\pm5$ & 255 & 0.03 & 1.47 & 1.0 & 101 & 0.55\\
6 & EV Lac & 2009-07-19 & $26\pm6$ & 1200 & 0.13 & 7.72 & 2.4 & 1326 & 7.0\\
7 & EV Lac & 2009-07-21 & $10\pm3$ & 300 & 0.04 & 0.391 & 0.4 & 24.7 & 0.13\\
8 & EV Lac & 2009-07-21 & \mbox{$14\pm2$}& 135 & 0.02 & 0.391 & 0.4 & 22.0 & 0.12\\
& & & $24\pm2$ & & & & & & \\
9 & Wolf 424 & 2010-06-10 & \mbox{$13\pm3$} & 320 & 0.07 & 1.75 & 1.1 & 120 & 0.040\\
& & & $57\pm17$ & & & & & & \\
10 & Wolf 424 & 2010-06-10 & $15\pm3$ & 459 & 0.07 & 3.85 & 1.7 & 76 & 0.025\\
11 & EV Lac & 2010-07-08 & $49\pm7$ & 910 & 0.03 & 0.248 & 0.2 & 69 & 0.38\\
12 & EV Lac & 2010-07-08 & $87\pm18$ &  &  &  &  &  & \\
13 & EV Lac & 2010-07-08 & $5.8_{-1.2}^{+0.9}$ & 340 & 0.03 & 0.426 & 0.4 & 30.2 & 0.16\\
14 & EV Lac & 2010-07-08 & \mbox{$18\pm7$} & 550 & 0.035 & 0.561 & 0.5 & 66 & 0.35\\
& & & $36\pm5$ & & & & & & \\
\hline
\hline
\end{tabular}
\label{T1}
\end{table*}

\begin{table}[hbt!]
\centering
\caption{Stellar effective temperatures and quiescent bolometric luminosities}
\begin{tabular}{c r r c  }
\hline
\hline
Star & $T_{\rm eff}$ & $L_{\rm bol}$ & reference \\
& (K) & ($L_{\rm Sun}$) & \\
\hline
UV Ceti & $2728 \pm 60$ & $(1.25 \pm 0.05) \times 10^{-3}$ & \citet{McD18}\\
Wolf 359& $2800 \pm 50$ & $(1.11 \pm 0.02) \times 10^{-3}$ & \citet{Cif20}\\
Wolf 424 & 3013 & $1.59 \times 10^{-3}$ & \citet{St13}\\
 & 2555/3800&  & \\
EV Lac & $3270 \pm 80$ & $(1.28 \pm 0.03) \times 10^{-2}$ & \citet{Pau21}\\
\hline
\hline
\end{tabular}
\label{T2}
\end{table}

\subsection{Energetics of flares and QPPs}

We prefer to characterize power of QPP using only the second QPP pulse, because the first pulses are standing out in the QPPs wavetrains and correspondingly don't represent the latters as is shown in Section~\ref{sec_Mor} below. The fractional flux amplitudes of the QPPs, $A_{\rm QPP} = {\rm \Delta} I/I_{\rm fl}$ for the second QPP pulse, and flares, $A_{\rm fl} = {\rm \Delta} I_{\rm fl}/I_{\rm q}$, are in the intervals 0.02--0.13 and 0.2--14.4, respectively, the latter corresponding to a stellar brightening of ${\rm \Delta}m_{\rm U} = $ 0.2--3.0 in the U filter (Table~\ref{T1}). The bolometric energy of a flare is estimated after \citet{Sh13} as $E_{\rm fl} = W_{\rm t} L_{\rm bol} \kappa_{\rm s}/4\kappa_{\rm fl} \eta$, where $W_{\rm t}$ is the equivalent duration of flare (ongoing with the quiescent stellar flux, we should also note its difference from the flare duration, ${\rm \Delta}t_{\rm fl}$), $L_{\rm bol}$ the quiescent star bolometric luminosity, $\kappa$ the fraction of the blackbody flux (determined by the effective temperature of star or flare, taken as a typical blackbody temperature of $9000$~K for a flare continuum \citep{HF92}) transmitted by the U filter, and $\eta = 0.63$ the fraction of the total flare-radiated energy in the optical continuum \citep{OW15}. With these we get the flares energies $E_{\rm fl}$ in the range (0.009--7.0)\,$\times 10^{32}$~erg (Table~\ref{T1}). The stellar effective temperatures and quiescent bolometric luminosities used in these estimations are given in Table~\ref{T2}.

\begin{figure*}
\centering
\includegraphics[width=1.0\linewidth]{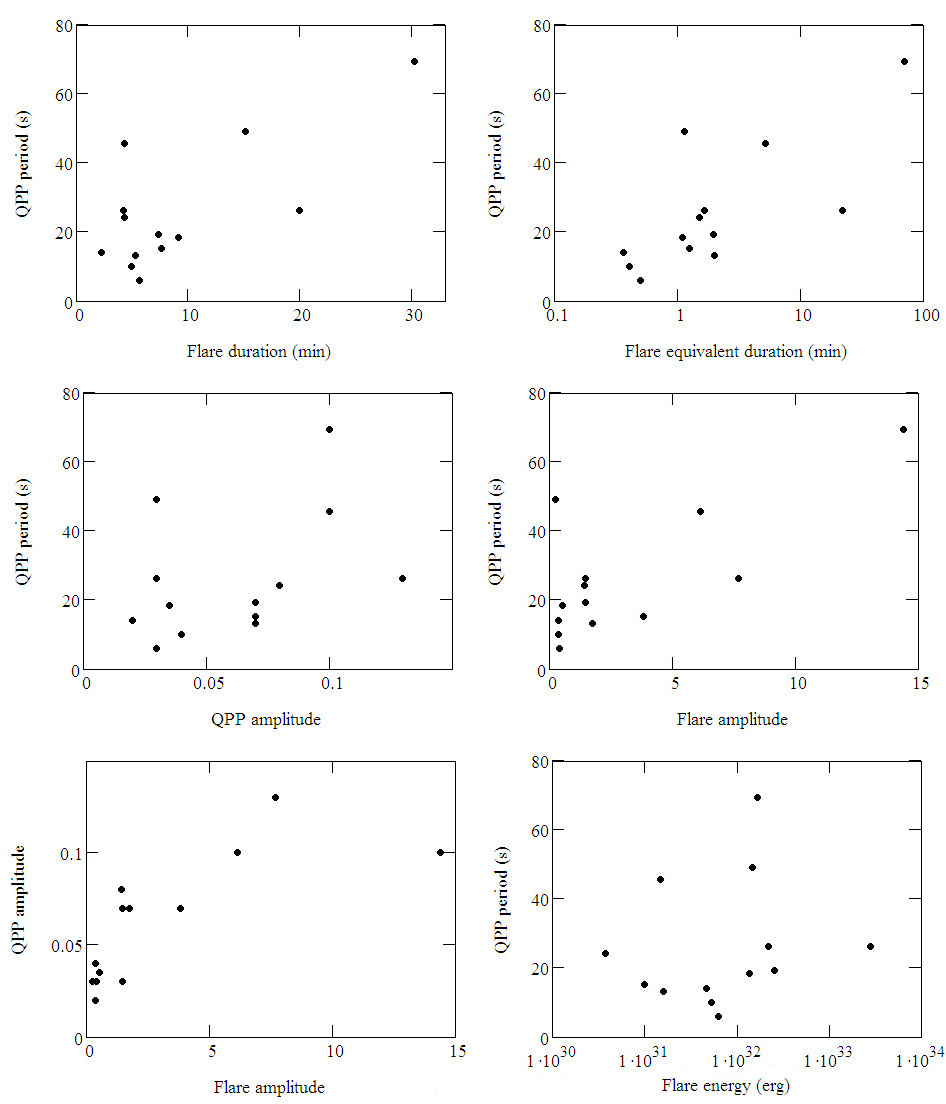}
\caption{Charts of the distributions of the characteristics of the M dwarfs QPPs in flares, which are indicative of the characteristics intercorrelations. The Pearson correlation coefficients and p-values are, respectively: 0.724 and 0.008 for the $P_{\rm QPP}$-${\rm \Delta}t_{\rm fl}$ correlation (the top-left chart), 0.731 and 0.008 for the $P_{\rm QPP}$-$W_{\rm t}$ correlation (the top-right chart), 0.412 and 0.087 for the $P_{\rm QPP}$-$A_{\rm QPP}$ correlation (the middle-left chart), 0.727 and 0.008 for the $P_{\rm QPP}$-$A_{\rm fl}$ correlation (the middle-right chart), 0.894 and $2\times 10^{-4}$ for the $A_{\rm QPP}$-$A_{\rm fl}$ correlation without the far right point (the bottom-left chart), 0.037 and 0.269 for the $P_{\rm QPP}$-$E_{\rm fl}$ correlation (the bottom-right chart).}
\label{P}
\end{figure*}

Some characteristics from Table~\ref{T1} are plotted in Fig.~\ref{P} against QPP period. It is evidently that the QPP period correlates with the duration (the Pearson correlation coefficient and p-value are, respectively, 0.724 and 0.008: hereafter the p-values lower than 0.05 statistically signifies a correlation), with the equivalent duration (0.731 and 0.008) and with the amplitude of a flare (0.727 and 0.008), but or weakly with the QPP amplitude (0.412 and 0.087). The absence of the correlation between the QPP period and the flare energy in the bottom-right panel of Fig.~\ref{P} (0.037 and 0.269) consists with the similar finding by \citet{Pug16} for the {\it Kepler} QPPs and by \citet{Hay20} for the Sun X-ray QPPs. This property suggests that the site of the flare energy release differs from the location of an oscillating medium. However, in energy release the site of the QPP emission scales with one of flare (0.894 and $2\times 10^{-4}$, without the far right point), as seen in the bottom-left panel of Fig.~\ref{P}, that could be if these sites coincide: then the size of the flare region would equally influence the both amplitudes, of flare and QPP. Moreover, there is noticeable even two branches of the scaling, though the high amplitude branch is scarcely represented.  

\subsection{Morphology of QPPs lightcurves}
\label{sec_Mor}

All found QPPs are very individual even for flares of the same star. Nevertheless there can be observed some common morphological features, namely:
\begin{itemize}
\item The first pulse in the QPP wavetrain, which occurs at flare peak. Usually it is either more intensive, preceding the weaker decayless wavetrain, or narrower, or more separated than following pulsations, or it has the combination of these properties. 
\item The spikes, a thin structure of the pulsations, occurring at the first pulse in particular.
\item The rapidly decaying first part of the QPP wavetrain where the decaying is going on during only 2--3 pulsations; more number of decaying pulses is an exception.
\item And the more prolonged decayless tail of the QPP wavetrain. Some QPPs consist only of this tail besides the first pulse, they have not the visible decaying first part. 
\item The suggestive traits on a superposition of pulsations: the saw-like profile of some pulses and the amplitude modulation of pulses in some decayless tails.
\end{itemize}
Now we consider these morphological features in individual QPPs which we group in the types of decaying QPPs, decayless QPPs and transitional QPPs (from decaying to decayless ones). Distinguishing these types we don't account for the first pulse because it seems standing out in the QPP wavetrains, in particular in the non-decaying QPPs, and it can mislead in the morphology characterization. To highlight a differentiation the QPPs in Fig.~\ref{PLCa} are grouped by the types: the decayings into the left column, the transitionals into the central column, and the decaylesses into the right column.  

\subsubsection*{Decaying QPPs}

{\em QPP~1}. This QPP belongs to the unique flare, described by \citet{Bes17}, where were discovered the subsecond spikes of synchrotron origin. The QPP morphology is entirely comprehended as a superposition of two pulsation, the relative narrowness of the first pulse in particular. This is consistent with the two pulsations revealed by the EMD analysis. The first pulse is spiked. \\
{\em QPP~12}. This and QPP~11 are the parts of single QPP, analysed separately from each other. Of note, the period $87 \pm 18$~s of the following part, QPP~12, could be a multiple of $49 \pm 7$~s, the period of the preceded QPP~11. The doubling of the QPP period was also mentioned by \citet{Doy22} for QPPs of M dwarf YZ~CMi\\
{\em QPP~14}. The first pulse, of the width 5~s, is essentially narrower than the rest and is badly reproduced by the EMD-modes. Such difference prompts to consider the first pulse separately from the rest. On the other hand, all together pulses could stem from the superposition as in the QPP~1 case, with the additive component of a $\sim 7$~s period which is present in the EMD spectrum under the significance level, possibly due to its shorter duration relatively the length of the light curve section under analysis. This suggestion of a hidden mode of the QPP consists with the finding by \citet{Doy22}, that among the QPP simultaneously initiated pulsations the shorter-period pulsation has shorter lifetime.

\subsubsection*{Decayless QPPs}

{\em QPP~8}. This QPP has two pulses besides the first pulse, which saw-like profile prompts the origin of the QPP morphology from the superposition of two pulsations as the EMD analysis shows. It is similar to the QPP~1 with the exception of the decaying tail, possible because of the shortening of the light curve in the QPP analysis. \\
{\em QPP~10}. Here the first pulse is highly intensive in comparison with the rest of this QPP. The decayless tail is amplitude modulated, that could be the trait of the superposition which short period mode is missed by the EMD analysis. This could also explain why the amplitudes of pulses are not reproduced by the EMD-modes. \\
{\em QPP~11}. Here the first pulse stands out among others, the EMD analysis doesn't render its narrowness and amplitude. In the tail of decayless pulsations there is the amplitude modulation which precedes the transition to the decaying {\em QPP~12} of approximately double period. \\ 
{\em QPP~13}. This is the clear case of the decayless amplitude modulated QPPs. The EMD analysis rejects the additional short period mode that is otherwise tempting because the pulses amplitudes are badly rendered. \\

\subsubsection*{QPP transitional between the decayings and decaylesses} 

{\em QPP~2}. Here the decaying goes on during 3 pulsations. The first two pulses are spiked and, that is noteworthy, have similar structure. Their amplitudes are not reproduced by the EMD-modes, that again suggests the superposition which short period component the EMD analysis missed. The 13~s pulsation of the decayless tail don't appear in the EMD spectrum. \\
{\em QPP~3}. This QPP is very similar to QPP~2 in both decaying and decayless parts and in badness of the EMD-modes for the pulse amplitudes, but the first pulse here is outstanding by narrowness. The proportions of widths and amplitudes of the first two pulses again could be explained by the superposition. \\
{\em QPP~4}. This looks like the QPP~2 that would be well gained. The amplitudes are not reproduced by the EMD-modes. \\
{\em QPP~5}. Here the flare is flat-toped. This QPP is also similar to QPP~2. The amplitudes are not reproduced by the EMD-modes. It seems that there is a $\sim 5$~s pulsation missed by the EMD model. \\
{\em QPPs~6, 7 and 9}. Here amplitudes of the decaying pulses are not reproduced by the EMD-modes. The decayless tails are amplitude modulated. 

It comes that the first pulse is robustly outstanding among others only in the QPP~11, and in the rest cases its peculiarity could be explained by the superposition of several pulsations some of which are missed by the EMD analysis.

\section{Conclusions}
\label{C}
Not a few flares of the nearest M dwarfs observed in a campaign of monitoring of microsecond variability in $U$-band showed QPPs. 13 flares of 44 appeared in QQPs that means 30\% occurrence of the optical QPPs among M dwarfs. The periods of these QPPs 6-107~s are beyond the periodicity range of the bulk known optical QPPs, mainly discovered by the observations of {\it Kepler} and TESS. This breaktrough exposition of QPPs is perhaps due to inherently high occurrence of QPPs of short periods, on the one hand, among weak stellar flares, on the other hand, which domains, of the QPPs periodicity and the flares energetics, were not vastly spanned by searchings for QPPs earlier. Incidentally, \citet{Hay20} don't exclude an observational bias in the detectability of QPPs among the solar X-ray flares, which could explain the smaller occurrence of QPPs among the (weaker) C-class flares than among the X-class flares. The energies of the flares revealing QPPs in our sample are mostly in the range $10^{30}$-$10^{32}$~erg.

At a rough guess, the found here QPPs could be morphologically grouped in the decaying QPPs, consisting only of 2-3 waves, the decayless QPPs, looking as a more extended wave train, and the QPPs transitional between them. The first pulse in a QPP, at flare maximum, is as rule distinguished from others. Therefore it is not accounted in this classification. In some cases it looks like a result of the superposition of several waves different in phase and period as in the case of QPP~1, where the superposition appears also in the saw-like profile of pulses. Such superposition is awaited for the stellar flares, which are not spatially resolved, where the source of QPPs is a bunch of oscillating coronal loops. The oscillations in the decayless parts of some QPPs are amplitude modulated, that also indicates the superposition. On the basis of this we suppose there are the three components which somehow constitute all QPPs: the first pulse detached from the following wave train, the next quickly decaying 2--3 waves and the following quasi-decayless pulsation, usually amplitude-modulated. Such QPP pattern looks like imprint in radiation of the sequence of magnetohydrodynamic wave processes similar to ones on the Sun observed in a flare by \citet{Nis14}, where a quasi-periodic fast ($> 1000$~km/s) quasi-decayless ($\ge 10$ wave fronts) wave trains propagate away from a shot of flare region. Indeed, the solar coronal loops display the evolution of kink oscillations from decaying to decayless state \citep{Nak24}, the behaviour we see in the stellar QPPs. 

The found QPPs display statistically significant correlations of the period with the duration, the equivalent duration and the amplitude of a flare. However, this should be considered tentatively because of the small number of the QPPs in the sample, besides non-uniformly distributed on period. The two, the correlation between the QPP amplitude and flare amplitude and the non-correlation between the QPP period and flare energy, seen at the bottom of Fig.~\ref{P}, consist with the scenarios of the modulation of the white light in a star flare by magnetohydrodynamic waves external to the emitting medium which is the source of the QPPs emission. 

\acknowledgements{
We would like to thank D.~Y.~Kolotkov for help with data analysis and for helpful discussions.
The work was partially carried out within the framework of the state assignment of the SAO RAS, approved by the Ministry of Science and Higher Education of the Russian Federation.
}

\end{document}